\documentclass[useAMS,usegraphicx,usenatbib]{mn2e}

\def\aap{{\sl A\&A}}
\def\aaps{{\sl A\&AS}}
\def\aj{{\sl AJ}}
\def\apj{{\sl ApJ}}
\def\apjl{{\sl ApJL}}
\def\apss{{\sl ApSS}}
\def\jrasc{{\sl JRASC}}
\def\mnras{{\sl MNRAS}}
\def\ocen{$\omega$ Cen}
\def\etal{{\sl et al}\ }
\def\msun{M_\odot}
\def\ltorder{<}

\title[A Simple Dynamical Model for \ocen]{A Simple Dynamical Model for \ocen}
\author[Mirek Giersz and Douglas C. Heggie]{Mirek Giersz$^{1}$\thanks{E-mail:
mig@camk.edu.pl (MG); d.c.heggie@ed.ac.uk (DCH)} and Douglas
C. Heggie$^{2}$
\footnotemark[1]
\\
$^{1}$Nicolas Copernicus Astronomical Center, Polish Academy of
Sciences, 00-716 Warsaw, ul. Bartycka 18, Poland\\
$^{2}$University of Edinburgh, School of Mathematics, King's
Buildings, Edinburgh EH9 3JZ, UK}
\begin{document}

\date{Accepted ??. Received ??; in original form 2002 August ??}

\pagerange{\pageref{firstpage}--\pageref{lastpage}} \pubyear{2002}

\maketitle

\label{firstpage}

\begin{abstract}
 We construct a simple dynamical model of the massive globular
cluster \ocen.  The model includes simple treatments of dynamical
evolution in a galactic tide, and evolution of single stars.  Binary
stars and rotation are neglected.  The model approximately fits
observational data on the surface brightness profile, the profile of
radial velocity dispersion, and the main sequence mass function at two
radii.
\end{abstract}


\section{Introduction}

Dynamical modelling of individual globular clusters has a long history
(see, for example, the review by Meylan \& Heggie 1997).  For the most
part the models used have been variants of the King-Michie model.
While incorporating essential aspects of stellar dynamics they have
nothing to say about dynamical {\sl evolution}, which has been the
focus of theory for many years.

Dynamical evolutionary models, based on Fokker-Planck codes, have been
constructed for a number of individual clusters by members of Cohn's
group (Grabhorn \etal 1992, Dull \etal 1997, Drukier 1995).  These
models were tested against available observational data on the surface
brightness profiles, individual stellar radial velocities, sometimes
ground-based mass functions, and even data on the exotic stellar
components in a cluster (Phinney 1993).

In the meantime a wealth of high quality data on the main sequence
mass function down to near the H-burning limit has been acquired,
thanks mainly to the high resolution of HST (e.g. Elson \etal 1995,
Cool \etal 1996, Santiago \etal 1996, von Hippel \etal 1996, Piotto
\etal 1997, King \etal 1998, Marconi \etal 1998, Pulone \etal 1999,
Paresce \& De Marchi 2000, De Marchi \etal 2000, Andreuzzi \etal 2001,
where we have restricted the credits to one citation per lead author).
Though static (King-like) models have been constructed which
incorporates some of this data (Anderson 1997, Sosin \& King 1997,
Saviane \etal 1998, Piotto \& Zoccali 1999, in addition to some of the
foregoing references), we are not aware of any evolutionary models
which do so.   

Our aim in this paper is a first step in this direction, i.e. to
construct a dynamical evolutionary model of a well observed, old,
galactic globular cluster, so as to represent its surface brightness
profile, profile of radial velocity dispersion, and mass function.
The object we have chosen for this project is \ocen\ (NGC 5139).  Being
massive, it has a long relaxation time, and is therefore dynamically
relatively unevolved.  Despite the uncertain effects of evolution of
stars above the turnoff, the mildness of any dynamical evolution makes
it relatively straightforward to guess appropriate initial conditions.
Nevertheless, some iteration is necessary, and for this purpose a fast
method of computing dynamical evolution is preferred.  We have adopted
a slightly modified version of a Monte Carlo code.

In the following section we first summarise the observational data we
have used.  Next we outline the manner in which the code was adapted
for this specific application, and how the output was converted for
comparison with observational data.  We then present our model, and
discuss briefly how the initial conditions were arrived at.  The final
section summarises our conclusions, and discusses the simplifying
assumptions on which our work is based.  

\section[]{A model of \ocen}

\subsection{Observational constraints}

\ocen\ is a famous object, and has even had its own conference
recently.  In the proceedings (van Leeuwen \etal 2002) our present
understanding of the many sides of \ocen\ is summarised in detail.  For
our purposes, however, the observational data will be restricted to
three sources: (i) a surface brightness profile; (ii) the dispersion
of radial velocities; and (iii) the mass function at two radii.

For the surface brightness profile we have
adopted the data in the compilation by Meylan (1987).  The surface
brightness profile consists
of both photometric data and values derived from star counts, the
latter having been normalised to fit the former in the overlap
region.  In modelling it might well be better to treat the two kinds
of data separately, but we have followed Meylan in treating the
surface brightness profile as a single data set.  The surface
brightness has been adjusted approximately for interstellar absorption.
Much doubt was cast on the reliability of this surface brightness
profile by van Leeuwen \etal (2000), on the basis of a proper motion
study, but more recently the results have been reconciled (van Leeuwen
\& Le Poole 2002).


Our radial velocity data is taken from Meylan \etal (1995).  We have
used the binned data presented in this paper, rather than individual
values in Meylan's catalogue.  This implies that the largest radius is
at about 28pc, compared with a nominal tidal radius of about 70pc
(Meylan 1987).  Therefore at large radii the surface brightness
profile is the only constraint.

For the mass function we have used the luminosity functions presented
by De Marchi (1999), converted to a mass function using the model of
Baraffe \etal (1997) for metallicity $[M/H] = -1.5$.  We have assumed
that the data of De Marchi give the numbers of observed stars per
magnitude bin in the two fields (with his stated adjustment by 0.1 dex
for the outer field).  This assumption is consistent with the stated
numbers of objects in each field.  The two fields for which mass
functions are given are stated to be at radii of about $7^\prime$ and
$4.6^\prime$, and these are the values we have adopted.  Nevertheless,
from an inspection of the image fields in relation to \ocen, we
consider that the outer field is centred more nearly at $7.4^\prime$.
We have also corrected these radii for the ellipticity, using values
from Geyer \etal (1983).  A more recent determination of the
ellipticity (van Leeuwen \& Le Poole 2002)
suggests that the ellipticity is even larger.

\subsection{ Features of the Monte Carlo code}

The Monte Carlo code (Giersz 1998, 2001) models a spherically symmetric
stellar system by a number of spherical shells characterised by
radius, energy and angular momentum.  As described in these papers the
number of such shells equals the number of stars in the real system.
In the present application, however, the number of shells is much
smaller than the number of stars, just as in the earlier formulations
by H\'enon (1971) and Stodo\l kiewicz (1982), and so
each shell is referred to as a {\sl superstar}.  Each superstar
represents a large number of stars of the same stellar mass, energy
and angular momentum. In this investigation the initial number of
shells was chosen in the range from $1024$ to $16384$.  Where
individual models are discussed below, $16384$ shells were used. 

Very briefly, the code chooses the radius of each star in a manner
appropriate to its energy and angular momentum.  The energy and
angular momentum are adjusted in a manner dictated by the theory of
two-body relaxation.  These processes are repeated until the required
evolution time (which we assume to be 12Gyr) has elapsed.

For dynamical purposes, stellar evolution is treated in a very simple
manner.  At the start, each star is assigned an evolution time
(depending on its initial mass)
according to the prescription adopted by Chernoff \& Weinberg (1990).
At this time the star is replaced by a degenerate remnant whose mass
is also determined as in Chernoff \& Weinberg, except for one point:
for an initial mass in the range $4.7\msun<m<8.0\msun$ the remnant is
assumed to be a neutron star of $1.4\msun$.  Neutron stars are given
no kick, and so all are retained, except those which may escape by
relaxation or tidal overflow; see also Sec.3.3.

Now we discuss briefly the data output from the Monte-Carlo code.  At
any time a model is defined by the radius, energy, angular momentum
and mass of its supershells.  The resulting surface brightness and
radial velocity dispersion profiles may be very noisy, for two
reasons.  First, each shell gives a cusp in surface brightness at its
edge.  Secondly, these profiles are dominated by the relatively small
number of evolving stars.  It is better, therefore, to represent each
superstar by a space density, corresponding to its orbital motion.  As
a compromise, we represented each superstar by 100 radii, chosen with
the correct radial distribution for a superstar with given energy and
angular momentum.

\subsection{Conversion to observational data}

Each shell of nominal radius $r$ is taken to represent a uniform
spherical shell of radii $0.9r$ and $1.1r$, to further reduce the
effects of the cusp which would exist at the projected edge of a thin
shell.  This shell also corresponds to known values of the radial and
transverse velocity.  It is therefore easy to compute a
density-weighted velocity dispersion along a line of sight.  Because
the observed radial velocities are obtained for giants, only shells
corresponding to non-degenerate stars above $0.7\msun$ were included.
(Though not all such stars would be giants, it was assumed that mass
segregation in the range of masses from $0.7\msun$ to the turnoff mass
could be ignored.)

The mass functions were obtained in a similar way.

The most problematic area is creation of the surface brightness
profile, which requires computation of the luminosity of each star
from its mass and age (12 Gyr).  For main sequence stars we used the
formulae of Eggleton \etal (1989), but scaled the evolution time
(i.e. the time for the end of non-degenerate phases of evolution) to coincide
with those used in the Monte Carlo code (i.e. those of Chernoff \&
Weinberg 1990).  

Applied to  evolving stars, this approach led to the occurrence of
a very few shells with very high luminosity, which produced a very
rough (``bumpy'') surface brightness profile.  Therefore we computed
the time-averaged luminosity during these phases of evolution, and
assigned this luminosity to each evolving star (i.e. post-main
sequence but non-degenerate stars).  For this purpose we used the code
of Hurley \etal (2000).  We checked that the mean luminosity of stars
brighter than $m_V = 16$ in the catalogue of Lyng\aa\ (1996) is approximately
consistent with what would be obtained from the code.

The surface brightness was corrected for extinction.
A simple bolometric correction was also applied (Reed 1998)\footnote{Note
that, in his formula (5), $T$ should be replaced by $10^{-4}T$}.

\section[]{Finding a model of \ocen}

\subsection{Initial conditions}

Our initial model is a King model (King 1966), specified by its total
mass, $M$, and scaled central potential, $W_0$.  The initial tidal
radius was set from observational estimates of {\sl current} values
($M = 3.9\times10^6M_\odot$ [Pryor \& Meylan 1993], $r_t = 63.9$pc,
from data in Trager \etal (1993) and Peterson (1993)), assuming that $r_t\propto
M^{1/3}$.

Guided by recent observational data (Kroupa 2001) we adopted an
initial mass function in the form of a continuous broken power law
$$
f(m) \propto \cases{m^{-\alpha_1}, & if $m_1<m<m_b$;\cr
		m^{-\alpha_2}, & if $m_b<m<m_2$.\cr}
$$
We quickly realised that $\alpha_1$ was quite tightly constrained near
$\alpha_1=1$ by the mass functions, and adopted this value.  We also
fixed $m_1=0.1\msun$ and $m_2=15\msun$.  The value of $m_1$ is a
little lower than the lowest mass included in the mass functions.
Specification of the upper mass limit $m_2$ is relatively unimportant
in our models: because the number of superstars is so modest, all
shells have masses considerably below $m_2$, unless $\alpha_2$ is
rather low.  The mass function is therefore specified by the mass at
the break point between the two power laws, $m_b$, and the slope of
the initial mass function for higher masses, $\alpha_2$.

In summary, each initial model is specified by the four parameters
$M$, $W_0$, $\alpha_2$, $m_b$.

\subsection{Exploration of initial conditions}

Using 4096 shells, the computation of a single model takes a few
minutes on a 400MHz Sun workstation.  After the Monte Carlo code has run,
the output can then be compared with the three kinds of data, i.e. the
profiles of surface brightness and velocity dispersion, and the mass
functions at two observed radii.  This
was done both visually and by a calculation of $\chi^2$.  For the
latter purpose, estimates of the errors of the observational data were
adopted.  Because of the Monte Carlo nature of the code, the
predictions of each model are also subject to statistical uncertainty,
but no attempt was made to quantify this for purposes of computation
of $\chi^2$. 

After preliminary examination of a number of models, the parameter
space was explored somewhat more systematically, but still manually,
by considering the effect of variation in each of the parameters.  The
conclusions are summarised in Table 1, which also gives the ranges of
values of the four parameters outside which the fit was observed to
deteriorate grossly (as judged both by graphical display and by the
values of $\chi^2$).  At this stage the best parameter values found
were approximately $M = 10^7\msun$, $W_0=8$, $\alpha_2=1.9$ and $m_b =
0.6\msun$.

\begin{table*}
\begin{minipage}{126mm}
 \caption{Preliminary parameter values.}
 \begin{tabular}{|l|l|l|}%
\hline
Parameter and range	&effect of increase
&effect of decrease\\
	&	&\\
\hline
$6\times10^6\msun<M<1.4\times10^7$ 	&mf and $v^2$ too
great	&mf and $v^2$ too low\\
$7.1<W_0<8.6$		&sfb too concentrated
&underluminous at centre\\
$1.7<\alpha_2<2.2$	&mf and sfb too high
&model underluminous; \\
		&	&central $v^2$ too small;\\
		&	&mf at inner radius poor\\
$0.6\msun<m_b<0.95\msun$	&sfb too concentrated,
	&centre overluminous, mf wrong\\
	&		&mf at smaller radius too low\\
\hline
 \end{tabular}
\medskip 
Notes: mf = mass function; sfb = surface brightness profile; $v^2$ =
radial velocity dispersion
\end{minipage}
\end{table*}

In order to improve these preliminary values several methods were
tried, and we describe here the two most successful ones.  The first
was simply to conduct a Monte Carlo search of the range constrained by
the values in Table 1, i.e. by uniform random sampling of the
corresponding hypercube in parameter space.  By plotting the resulting values of $\chi^2$
against each parameter, it was quite easy to determine the best values
with acceptable accuracy.  This is a relatively slow method, however,
and requires thousands of Monte Carlo runs.  A faster and automatic
method, requiring only of order 50 runs, treats our problem as one of
stochastic optimization, the stochasticity arising from the nature of
the Monte Carlo method used for the dynamical evolution.  Known simply
as DIRECT, for ``DIviding RECTangles'' (Jones 2001), it proceeds by
subdividing the search domain in a manner that balances global and
local searches for a minimum.


\subsection{Features of a typical model}

Both methods described in the previous subsection led to fairly
consistent conclusions.  The first method yielded initial conditions $M\simeq 1.0\times
10^7M_\odot$, $W_0 \simeq7.7$, $\alpha_2\simeq1.9$ and
$m_b\simeq0.6M_\odot$, while DIRECT gave results in the ranges $0.94\times10^7\msun\ltorder
M\ltorder1.23\times10^7\msun$, $7.4\ltorder W_0\ltorder 7.9$,
$1.95\ltorder\alpha_2\ltorder2.12$ and $0.63\msun\ltorder
m_b\ltorder1.14\msun$.  One of the best models is illustrated in
Fig.1.  The current mass is $3.6\times10^6\msun$, which is perhaps a
little too small: the resulting tidal radius is perhaps a little too
small to account for the surface brightness profile at the largest
radii.  On the other hand the modelling of the tide as a cutoff is
inaccurate near the tidal radius, and so a good fit here may not be
achievable. 

We have not tried to adjust the initial mass function on the lower
main sequence to improve the detailed fit with the mass function.
Of greater concern is the fact that the model mass functions are often
slightly but systematically too low or high, at the inner and outer
radii, respectively.  The suggestion that the ellipticity exceeds the
value we used (see Sec.2.1) would help.

In attempting to construct a multi-mass King model for \ocen, Meylan
(1987) drew attention to the need for heavy remnants, by which
we mean here both neutron stars and white dwarfs.  Our models include such
remnants, which arise from the evolution of stars above the turnoff
mass in our mass function.  Their proportion by mass at the present
day, in our best models, is of order 50\%, and even so it is not
possible to quite reach the observed velocity dispersion at small
radii.  This problem worsens if all neutron stars are removed at birth
(i.e. it is assumed that each receives a kick exceeding the escape
speed).  Though the total fraction of degenerate stars declines only to
about 40\%, the central velocity dispersion drops to about $12$km/s.


\begin{figure*}
\begin{minipage}{126mm}
\vspace{2.7truein}
\includegraphics{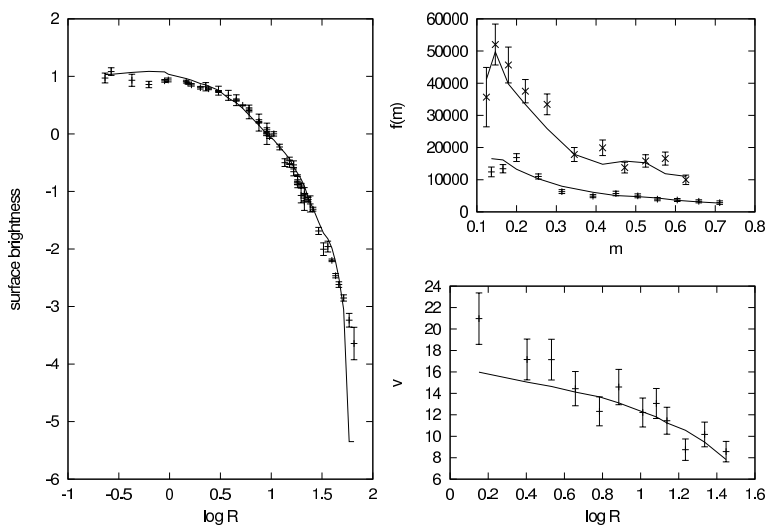}
\caption{One of the best models.  Here $M = 0.94\times10^7\msun$, $W_0 =
7.6$, $\alpha_2 = 1.95$ and $m_b = 0.63\msun$; the number of
superstars was $16384$.  Left: logarithmic surface
brightness profile (in units of 10.00 $V$ mag per square arc minute)
against $R$ (pc); lower right: velocity dispersion profile (km/s);
upper right: mass function (stars per unit mass per square arc minute;
the upper and lower plots correspond to the inner and outer observed
radii, respectively).}

\end{minipage}
\end{figure*}

Several other features of these models may be of interest.  The models
are
mildly anisotropic.  If the anisotropy parameter $\beta$ is defined by
$\beta = 1- \langle v_r^2\rangle/\langle v_t^2\rangle$, where $v_r$,
$v_t$ are the radial and transverse velocities {\sl in the plane of
the sky} (i.e. as measured by proper motions), we find that $\beta$
varies from a value of about 0 within the innermost 2pc to about
$-0.15$ at a radius of 10 pc, and then rises towards 0 as the tidal
boundary is approached.  At 20 pc, close to the radius where King \&
Anderson (2002) found only mild anisotropy, $\beta\simeq-0.05$.


The models also exhibit mild mass segregation.  In the model exhibited
in Fig.1, the mean mass of unevolved stars is nearly $0.40M_\odot$ at
all (projected) radii less than about 1pc, and about 0.34 beyond 10pc;
the mean mass declines steadily between 1 and 10pc.  The primordial
value (over the same range of stellar masses) was $0.35M_\odot$.  For
a single power law $f(m)dm \propto m^{-\alpha} dm$ between the minimum
mass and turnoff, the variation of mean mass corresponds to a
variation of $\alpha$ of about 0.5.  Relative to the centre,
therefore, there is an excess of stars of lowest mass at the outside
of the cluster by a factor of order 3 (0.5dex).  Anderson (2002) has observed
slight mass segregation by comparing the luminosity function at the
centre relative to a field at about $7^\prime$.  His faintest stars
are underabundant at the centre by about 0.2dex, but do not extend to
such low masses.

The fact that the signatures of dynamical evolution in \ocen\
are not great is no surprise, but this is perhaps the first time they
have been quantified theoretically.

\section[]{Discussion and conclusions}

\subsection{Discussion}

Before summarising the tentative findings of this study, it is
important to set out some aspects of \ocen\ which we have not included.

The most important constituent we have omitted is any population of
binary stars.  At the expense of introducing further parameters, it
would have been possible to do so by treating the population as
dynamically ``inert'', i.e. as simply a population of slightly more
massive stellar objects.  Nevertheless it would have been desirable to
treat their stellar evolution in a fundamentally different way from
that of the single stars, and so for this exploratory study they were
neglected entirely.  If included, they might have helped to deepen the
potential well and increase the central velocity dispersion, without
unduly distorting the surface brightness there.  They might also
assist the retention of neutron stars, with the same result.  Our treatment of the
evolution of single stars could also be improved significantly.

Dynamically, we have taken no notice of the fact that \ocen\ is
rotating (see Merritt \etal 1997).  For purposes of dynamical evolution it
seems that rotation does not play a dominant role (Spurzem 2001), but
our practical reason for omitting rotation is that the code cannot
cope with it.  The rotation of \ocen\ may be a symptom of past mergers
(Norris \etal 1997), and this possibility is ignored here also.

In fitting models to observational data we have ignored the
kinematical evidence from internal proper motions (van Leeuwen \etal
2000).  This data appears to be entirely consistent with the radial
velocity, which we did employ, and also covers a similar range of
radius within the cluster.  Another reason for neglecting
this data is that we have it in mind to apply our methods to several
other clusters for which such data is lacking entirely. 

A significant dynamical mechanism that we have also ignored is the
time taken for stars to escape.  As Baumgardt (2001) has shown, the
effect is that the lifetime in a tidal field is not proportional to
the relaxation time, as we would find using our Monte Carlo code.  On
the other hand we have also simplified the treatment of the tidal
boundary condition by supposing that the tide is steady, as for a
cluster on a circular galactic orbit.  We have made no attempt to
model the tidal debris of \ocen, whose mass is considerable (Leon
\etal 2000).

An issue which we would have liked to address is the uniqueness of the
model initial conditions that we have found.  While we have explored
the parameter space in various ways, and have tried to place bounds
on the parameters which give acceptable models, perhaps radically
different models are possible.  

Our best model is no more than a tolerable fit to the data with which
it has been compared.  On the other hand the data itself is not without
problems, such as the difficulty of converting from a magnitude
distribution to a mass function.

\subsection{Conclusions}

We have found that the surface brightness profile, velocity dispersion
profile and mass function of \ocen\ can be fitted approximately by the
dynamical evolution, over 12Gyr, of a cluster with the following
initial conditions:  the initial mass is about $1.1\times10^7M_\odot$, the initial tidal
radius is about $90$pc, and the initial model is a King model with a
scaled central potential $W_0 = 7.7$ approximately;  the initial mass function
is a broken power law, with slopes of about $1$ and $1.9$ respectively below and above
the break-point mass of about $0.6\msun$.  

The resulting present mass
and tidal radius are about $3.6\times10^6M_\odot$ and 61pc, respectively.  The current
proportion of mass
in heavy remnants in our model is about 55\%.

\section*{Acknowledgments}

DCH thanks CAMK for its cheerful hospitality during a visit when this
research was begun, and MG similarly thanks Edinburgh for its
hospitality during a visit, supported under grant 2 PO3D 024 19 of the
Polish National Committee for Scientific Research, when it was largely
completed.  We are grateful to R. Butler and J. Kaluzny for their
advice on observational matters, and J.A.J. Hall for pointing us in
the direction of DIRECT.  We thank the referee also for his attentive
comments.  We gratefully acknowledge the use we
have made of the internet facilities of the HST ECF, CDS and ADS.

\bsp

\label{lastpage}


\begin{thebibliography}{}
\bibitem[\protect\citeauthoryear{}{}]{1997PhDT.........8A}Anderson, A.~J.\ 1997, 
Ph.D.~Thesis, Univ. of California, Berkeley
\bibitem[\protect\citeauthoryear{}{}]{}Anderson, A.~J.\ 2002, in van Leeuwen F.,
Hughes J.,  Piotto G., eds, $\omega$ Centauri -- A Unique Window into
Astrophysics, ASP Conf. Ser. 265, ASP, San Francisco, p.87
\bibitem[\protect\citeauthoryear{}{}]{2001A&A...372..851A}Andreuzzi, G., De 
Marchi, G., Ferraro, F.~R., Paresce, F., Pulone, L., \& Buonanno, R.\ 2001, 
\aap, 372, 851 
\bibitem[\protect\citeauthoryear{}{}]{1997A&A...327.1054B} Baraffe, I., Chabrier, G., Allard, 
F., \& Hauschildt, P.~H.\ 1997, \aap, 327, 1054 
\bibitem[\protect\citeauthoryear{}{}]{1997A&A...327.1054B} Baumgardt,
H.,\ 2001, \mnras, 325, 1323
\bibitem[\protect\citeauthoryear{}{}]{1990ApJ...351..121C} Chernoff, 
D.~F.~\& Weinberg, M.~D.\ 1990, \apj, 351, 121 
\bibitem[\protect\citeauthoryear{}{}]{1996ApJ...468..655C} Cool, A.~M., 
Piotto, G., \& King, I.~R.\ 1996, \apj, 468, 655 
\bibitem[\protect\citeauthoryear{}{}]{1999AJ....117..303D} De Marchi, G.\ 1999, \aj, 
117, 303 
\bibitem[\protect\citeauthoryear{}{}]{2000ApJ...530..342D} De 
Marchi, G., Paresce, F., \& Pulone, L.\ 2000, \apj, 530, 342 
\bibitem[\protect\citeauthoryear{}{}]{}  Drukier G.A., 1995, {\sl ApJS}, {\bf 100}, 347
\bibitem[\protect\citeauthoryear{}{}]{}  Dull J.D., Cohn H.N., Lugger P.M., Murphy B.W.,
Seitzer P.O., Callanan P.J., Rutten R.G.M., Charles P.A., 1997, {\sl
ApJ}, {\bf 481}, 267
\bibitem[\protect\citeauthoryear{}{}]{1989ApJ...347..998E} Eggleton, 
P.~P., Tout, C.~A., \& Fitchett, M.~J.\ 1989, \apj, 347, 998 
\bibitem[\protect\citeauthoryear{}{}]{}  Elson R.A.W., Gilmore G.F., Santiago B.X.,
 Casertano S., 1995, {\sl AJ}, {\bf 110} 682
\bibitem[\protect\citeauthoryear{}{}]{1983A&A...125..359G} Geyer E.~H., Nelles B., 
Hopp U., 1983, \aap,  125, 359 
\bibitem[\protect\citeauthoryear{}{}]{1998MNRAS.298.1239G} Giersz, M.\ 1998, \mnras, 298, 
1239 
\bibitem[\protect\citeauthoryear{}{}]{2001MNRAS.324..218G} Giersz, M.\ 2001, \mnras, 324, 
218 
\bibitem[\protect\citeauthoryear{}{}]{}  Grabhorn R.P., Cohn H.N., Lugger P.M., Murphy B.W., 1992,
{\sl ApJ}, {\bf 392}, 86
\bibitem[\protect\citeauthoryear{}{}]{1971Ap&SS..14..151H} H{\' e}non, M.~H.\ 1971, 
\apss, 14, 151 
\bibitem[\protect\citeauthoryear{}{}]{2000MNRAS.315..543H} Hurley, J.~R., 
Pols, O.~R., \& Tout, C.~A.\ 2000, \mnras, 315, 543 
\bibitem[\protect\citeauthoryear{}{}]{}Jones D.R., 2001, in Floudas, C.A., Pardalos, P., eds,
Encyclopaedia of Optimization, Kluwer, Dordrecht, p.431;  see also
http://www4.ncsu.edu/~ctk/iffco.html 
\bibitem[\protect\citeauthoryear{}{}]{1966AJ.....71...64K} King I.~R., 1966, \aj,  71, 64 
\bibitem[\protect\citeauthoryear{}{}]{} King I.R., Anderson J., 2002, in van Leeuwen F.,
Hughes J.,  Piotto G., eds, $\omega$ Centauri -- A Unique Window into
Astrophysics, ASP Conf. Ser. 265, ASP, San Francisco, p.21
\bibitem[\protect\citeauthoryear{}{}]{1998ApJ...492L..37K} King, 
I.~R., Anderson, J., Cool, A.~M., \& Piotto, G.\ 1998, \apjl, 492, L37 
\bibitem[\protect\citeauthoryear{}{}]{2001dscm.conf..187K} Kroupa P.,
2001, in Deiters S., Fuchs B., Just A., Spurzem R., Wielen R., eds,
Dynamics of Star Clusters and the Milky Way, ASP Conf. Ser. 228, ASP, San Francisco, p.187
\bibitem[\protect\citeauthoryear{}{}]{2000A&A...359..907L} Leon S., Meylan G., Combes 
F., 2000, \aap,  359, 907 
\bibitem[\protect\citeauthoryear{}{}]{1996A&AS..115..297L} Lyng\aa, G.\ 1996, \aaps, 115, 297 
\bibitem[\protect\citeauthoryear{}{}]{1998MNRAS.293..479M} Marconi, G., Buonanno, R., Carretta, E., Ferraro, F. R., Montegriffo, P.,
 Fusi Pecci, F., De Marchi, G., Paresce, F.,
 Laget, M.,\ 1998, \mnras, 293, 479 
\bibitem[\protect\citeauthoryear{}{}]{1997AJ....114.1074M} Merritt D., Meylan G., 
Mayor M., 1997, \aj,  114, 1074 
\bibitem[\protect\citeauthoryear{}{}]{1987A&A...184..144M} Meylan, G.\ 1987, \aap, 184, 144 
\bibitem[\protect\citeauthoryear{}{}]{} Meylan G., Heggie D.C., 1997, {\sl A\&A Rev}, {\bf 8}, 1 
\bibitem[\protect\citeauthoryear{}{}]{1995A&A...303..761M} 
Meylan, G., Mayor, M., Duquennoy, A., \& Dubath, P.\ 1995, \aap, 303, 761 
\bibitem[\protect\citeauthoryear{}{}]{1997ApJ...487L.187N} Norris J.~E., Freeman 
K.~C., Mayor M., Seitzer P., 1997, \apjl,  487, L187 
\bibitem[\protect\citeauthoryear{}{}]{2000ApJ...534..870P} Paresce, F.~\& De 
Marchi, G.\ 2000, \apj, 534, 870 
\bibitem[\protect\citeauthoryear{}{}]{1993sdgc.proc..337P} Peterson
C.~J., 1993, in Djorgovski, S.G., Meylan, G., eds, Structure and
Dynamics of Globular Clusters, ASP Conf. Ser. 50, ASP, San Francisco, p.337 
\bibitem[\protect\citeauthoryear{}{}]{} Phinney E.S., 1993, in {\sl Structure and Dynamics of
Globular Clusters}, eds Djorgovski S.G., Meylan G., ASP
Conf. Ser. {\bf 50}, (ASP, San Francisco)
\bibitem[\protect\citeauthoryear{}{}]{1997AJ....113.1345P} Piotto, G., 
Cool, A.~M., \& King, I.~R.\ 1997, \aj, 113, 1345 
\bibitem[\protect\citeauthoryear{}{}]{1999A&A...345..485P} Piotto,
G.~\& Zoccali, M.\ 1999, \aap, 345, 485
\bibitem[\protect\citeauthoryear{}{}]{1993sdgc.proc..357P} Pryor C., Meylan G., 
1993, in Djorgovski, S.G., Meylan, G., eds, Structure and
Dynamics of Globular Clusters, ASP Conf. Ser. 50, ASP, San Francisco, p.357
\bibitem[\protect\citeauthoryear{}{}]{1999A&A...342..440P} Pulone, 
L., De Marchi, G., \& Paresce, F.\ 1999, \aap, 342, 440 
\bibitem[\protect\citeauthoryear{}{}]{1998JRASC..92...36R} Reed B.~C., 1998, \jrasc,  92, 36 
\bibitem[\protect\citeauthoryear{}{}]{}  Santiago B.X., Elson R.A.W., Gilmore G.F., 1996, {\sl MNRAS},
{\bf 281}, 1363
\bibitem[\protect\citeauthoryear{}{}]{1998A&A...333..479S} Saviane, I., Piotto, 
G., Fagotto, F., Zaggia, S., Capaccioli, M., \& Aparicio, A.\ 1998, \aap, 
333, 479 
\bibitem[\protect\citeauthoryear{}{}]{1997AJ....113.1328S} Sosin, C.~\& King, 
I.~R.\ 1997, \aj, 113, 1328 
\bibitem[\protect\citeauthoryear{}{}]{} Spurzem, R.\ 2001, in  Deiters S., Fuchs B., Just A., Spurzem R., Wielen R., eds,
Dynamics of Star Clusters and the Milky Way, ASP Conf. Ser. 228, p.75
\bibitem[\protect\citeauthoryear{}{}]{1982AcA....32...63S} Stodo\l kiewicz, J.~S.\ 
1982, Acta Astronomica, 32, 63 
\bibitem[\protect\citeauthoryear{}{}]{1993sdgc.proc..347T} Trager S.~C., Djorgovski 
S., King I.~R., 1993, in Djorgovski, S.G., Meylan, G., eds, Structure and
Dynamics of Globular Clusters, ASP Conf. Ser. 50, ASP, San Francisco, p.347 
\bibitem[\protect\citeauthoryear{}{}]{2000A&A...360..472V} van Leeuwen, F., Le 
Poole, R.~S., Reijns, R.~A., Freeman, K.~C., \& de Zeeuw, P.~T.\ 2000, 
\aap, 360, 472 
\bibitem[\protect\citeauthoryear{}{}]{} van Leeuwen F.,  Hughes J.,  Piotto G., eds, 2002, {\sl
Omega Centauri -- a unique window into astrophysics}, ASP
Conf. Ser. 265, ASP, San Francisco
\bibitem[\protect\citeauthoryear{}{}]{} van Leeuwen F., Le Poole R.S., 2002, in van Leeuwen F.,
Hughes J.,  Piotto G., eds, $\omega$ Centauri -- A Unique Window into
Astrophysics, ASP Conf. Ser. 265, ASP, San Francisco, p.41
\bibitem[\protect\citeauthoryear{}{}]{}  von Hippel T., Gilmore G., Tanvir N., Robinson D.,
 Jones D.H.P., 1996, {\sl AJ}, {\bf 112}, 192
\end{thebibliography}
\end{document}